\pdfoutput=1
\documentclass[aps,twocolumn,showpacs,superscriptaddress,a4paper,floatfix,longbibliography]{revtex4-1}

\usepackage[utf8]{inputenc}
\usepackage{siunitx}
\usepackage{hyperref}
\usepackage{datetime}
\usepackage{graphicx}
\usepackage[dvipsnames]{xcolor}
\usepackage{amsmath,amssymb}

\usepackage{textcomp}

\DeclareSIUnit\gauss{G}

\hypersetup{
    colorlinks,
    linkcolor={red!50!black},
    citecolor={blue!50!black},
    urlcolor={blue!80!black}
}

\newcommand{\extref}[2][]{\hyperref[#2]{\ref*{#2}#1}}

\begin{document}

\title{Interaction-Assisted Reversal of Thermopower with Ultracold Atoms}

\author{Samuel Häusler}
\author{Philipp Fabritius}
\author{Jeffrey Mohan}
\affiliation{Department of Physics, ETH Zurich, 8093 Zurich, Switzerland}
\author{Martin Lebrat}
\affiliation{Department of Physics, Harvard University, Cambridge, Massachusetts 02138, USA}
\author{Laura Corman}
\email{lauracorman@xrite.com}
\altaffiliation[Present address: ]{X-Rite Europe GmbH, 8150 Regensdorf, Switzerland}
\author{Tilman Esslinger}
\email{esslinger@phys.ethz.ch}
\affiliation{Department of Physics, ETH Zurich, 8093 Zurich, Switzerland}

\date{\pdfdate}

\begin{abstract}
We study thermoelectric currents of neutral, fermionic atoms flowing through a mesoscopic channel connecting a hot and a cold reservoir across the superfluid transition. The thermoelectric response results from a competition between density-driven diffusion from the cold to the hot reservoir and the channel favoring transport of energetic particles from hot to cold. We control the relative strength of both contributions to the thermoelectric response using an external optical potential in a nearly non-interacting and a strongly-interacting system. Without interactions, the magnitude of the particle current can be tuned over a broad range but is restricted to flow from hot to cold in our parameter regime. Strikingly, strong interparticle interactions additionally reverse the direction of the current. We quantitatively model \emph{ab initio} the non-interacting observations and qualitatively explain the interaction-assisted reversal by the reduction of entropy transport due to pairing correlations. Our work paves the way to studying the coupling of spin and heat in strongly correlated matter using spin-dependent optical techniques with cold atoms.
\end{abstract}

\maketitle

\section{Introduction}

Transport of charge, heat, and spin are often coupled in nature. Their interplay enriches the dynamical response of materials leading to coupled transport phenomena such as thermoelectricity \cite{He2017} and, along conceptually similar lines, spintronics \cite{Zutic2004} and spin caloritronics \cite{Bauer2012}. Thermoelectricity includes two major observations where an applied temperature gradient can induce a charge current (Seebeck effect) or an external voltage can give rise to a heat current (Peltier effect). Besides their widespread practical applications, both are essential to probe fundamental physics. In particular, studies of these effects in strongly correlated materials have allowed researchers to identify relevant charge carriers \cite{Tomczak2018, Cyr-Choiniere2017, Laliberte2011} and degrees of freedom \cite{Wang2003}, which have been proposed to characterize exotic states such as Majorana modes or anyons \cite{Kleeorin2019}.

Microscopically, thermoelectric currents in conventional materials originate from an electron-hole asymmetry created by an energy-dependent density of states and carrier velocity. In the case of a temperature gradient, for instance, this asymmetry can favor the transport of high-energy particles from the hot side over the transport of low-energy particles from the cold side. This imbalance results in a net carrier flow whose magnitude increases with asymmetry.

In solid-state systems, several techniques have been explored to engineer the thermoelectric response. First, the energy dependence of the density of states can be enhanced by reducing the number of free dimensions \cite{Dresselhaus1999, Dresselhaus2007} and using electrostatic gate potentials in low-dimensional structures such as quantum wires \cite{Dubi2011}, point contacts \cite{Houten1992}, and dots \cite{Staring1993}. Furthermore, the thermoelectric response can be strongly modified by electron interactions as observed in quantum dots \cite{Staring1993} and two-dimensional electron gases \cite{Mokashi2012}. However, the interpretation of the thermoelectric response in solids is complicated by interactions of the carriers with impurities, defects, and phonons.

Because of the absence of these factors, ultracold atoms are well suited to simulate the relevant physics of real materials. In addition, Feshbach resonances allow one to study the same system across a large range of interaction strengths under comparable conditions. These merits have facilitated experiments on transport phenomena in strongly interacting Fermi gases, including viscous flow \cite{Cao2011}, spin diffusion \cite{Sommer2011, Koschorreck2013}, sound propagation \cite{Patel2020, Bohlen2020}, and heat transport in the form of second sound \cite{Sidorenkov2013}.

As these experiments focused on bulk material properties, they lacked the tunability of mesoscopic systems; however, recent work in optically shaping bulk gases made it possible to create mesoscopic cold atom ``devices'' comparable to their solid-state counterparts \cite{Albiez2005, Brantut2012, Eckel2014, Jendrzejewski2014, Krinner2015, Valtolina2015, Chien2015, Krinner2017, Amico2020}. In particular, thermoelectric phenomena were explored in these mesoscopic structures, focusing on either weak \cite{Brantut2013} or strong \cite{Husmann2018} interactions. Here, we exploit the ability to compare both interaction regimes in the same structure and, by extending the accessible range of gate potentials, observe a striking reversal of the thermoelectric current directly induced by interactions. This reversal is a novel effect in cold atoms and, to our knowledge, also in strongly correlated solid materials. With the gate, we can finely tune the particle-hole asymmetry at the origin of the thermoelectric current and thereby engineer both its magnitude and direction. We can therefore smoothly turn our system from a heat engine into a heat pump, the latter being an important ingredient for an efficient cooling scheme proposed for cold atoms \cite{Grenier2014}.

For weak interactions, the thermoelectric response can be predicted by an \emph{ab initio} Landauer model thanks to the absence of defects and precise characterization of our system. With strong interactions, we focus on temperatures around the superfluid transition where a large critical region is predicted \cite{Debelhoir2016} and Fermi liquid behavior breaks down \cite{Sagi2015}. Although the strongly correlated regime is often challenging to understand, we interpret our observation on a fairly fundamental level based on entropy transport. Other works have studied theoretically thermoelectric effects with cold atoms for bosonic \cite{Karpiuk2012, Papoular2014, Rancon2014, Filippone2016, Uchino2020_0} and fermionic \cite{Sekera2016, Pershoguba2019_0, Uchino2020} systems.

The structure of the paper is as follows. After introducing the setup in Sec.~\ref{sec:setup}, we explain its thermoelectric response with an intuitive picture [Sec.~\ref{sec:landauer_picture}] and discuss the dynamics in the non- and strongly-interacting regimes [Sec.~\ref{sec:dynamics}]. Based on a phenomenological model presented in Sec.~\ref{sec:phenomenological_model}, we extract transport properties and discuss their behaviors in Secs.~\ref{sec:thermopower} and \ref{sec:lorenz_number}. Finally, we conclude in Sec.~\ref{sec:discussion}. Technical details can be found in the Appendixes.

\section{Setup}
\label{sec:setup}

Our transport setup consists of a mesoscopic channel smoothly connected to reservoirs of degenerate fermions ($^6$Li), as sketched in Fig.~\extref[(a)]{fig:concept} and described in previous works \cite{Stadler2012, Brantut2012}. The channel is created by a repulsive TEM$_{01}$-like laser beam that confines the atoms along the $z$ direction to its dark nodal plane, reaching a trapping frequency $\nu_{cz} = \SI{4.5(8)}{\kilo\hertz}$ at the center. Its Gaussian envelope in the longitudinal direction (\(y\)) ensures the smooth connection of the channel and reservoirs. In the $x$ direction, the atoms are restricted by the dipole trap, providing a weak confinement with frequency $\nu_{tx} = \SI{232(1)}{\hertz}$.

The reservoirs, denoted by left (L) and right (R), contain equal atom numbers, with $N = N_L + N_R = \num{121(2)e3}$ atoms in each of the two lowest hyperfine states. To prepare a temperature difference, we heat one side by parametrically modulating the intensity of an attractive beam inside one reservoir while blocking the channel. Subsequent equilibration of each reservoir leads to a temperature difference $\Delta T = T_L - T_R = \SI{147(11)}{\nano\kelvin}$ and an average value $\bar{T} = (T_L + T_R) / 2 = \SI{208(6)}{\nano\kelvin}$. After reconnecting the reservoirs, they exchange particles and heat through transverse modes of the channel. Their energy is controlled by one of the two gate beams, where one is repulsive ($V_g > 0$) and the other attractive ($V_g < 0$) [Fig.~\extref[(a)]{fig:concept}]. By tuning the gate, the number of available modes below the chemical potential $\bar{\mu} = (\mu_L + \mu_R) / 2 = \SI{151(16)}{\nano\kelvin}$ reaches up to 40 with 4 occupied states in the tightly confined $z$ direction making the channel quasi--two dimensional \footnote{In our case, the confinement does not induce pairing even in the unitary limit. There, the binding energy in a two-dimensional system is $0.232 h \nu_{cz} \sim \SI{50}{\nano\kelvin}$, which is small compared to our temperature of $\bar{T} \sim \SI{200}{\nano\kelvin}$ \cite{Sommer2012, Bloch2008}.}. Note that non-zero temperature leads to a partial occupation of modes at energies above the chemical potential, which also contribute to transport.

Using a broad Feshbach resonance of $^6$Li, we set the interparticle interactions in the entire system, including the channel and the reservoirs, either close to zero (non-interacting) or on resonance (unitarity). Subsequently, we enable transport for a variable time $t$ and measure the differences in atom numbers $\Delta N (t)$ and temperature $\Delta T (t)$ from an absorption image. The initial conditions stated above correspond to the non-interacting case and are given for the unitary case in Appendix~\ref{app_sec:experimental_details}, together with additional experimental details.

\begin{figure}
    \includegraphics[width=3.3in]{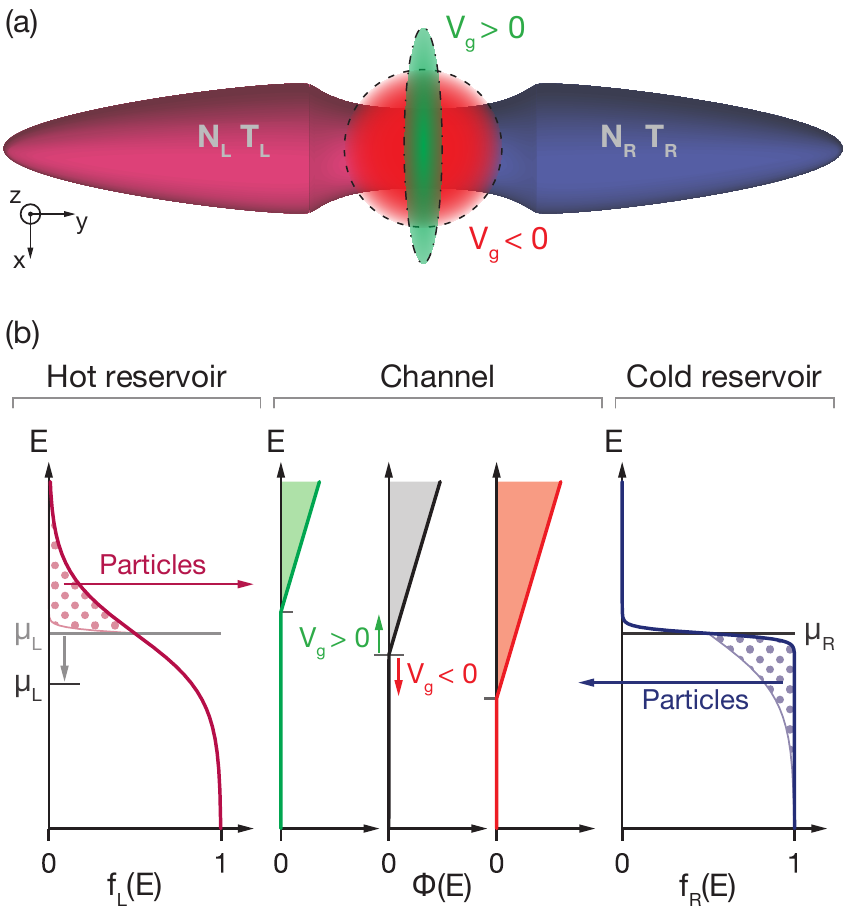}
    \caption{Concept and experimental setup. (a) Schematic view of a quasi-two-dimensional channel connected to a hot (red) and a cold (blue) reservoir of fermionic atoms. Their unequal temperatures $T_L$ and $T_R$ induce a net particle current that changes the initially equal atom numbers $N_L$ and $N_R$ over time. The thermoelectric response is tuned by a gate potential $V_g$ inside the channel that is either attractive (red) or repulsive (green).
    (b) Roles of the reservoirs and channel in thermoelectric transport. The reservoirs inject particles into the channel according to their occupations described by the Fermi-Dirac distributions $f_L (E)$ and $f_R (E)$. As the hot (cold) side contains more particles at high (low) energies compared to the other side (dotted regions), two counter-propagating currents at different energies emerge (horizontal arrows). At equal atom numbers, the chemical potential $\mu_L$ in the hot reservoir is lower than $\mu_R$ on the cold side, introducing an asymmetry between them. A particle at energy $E$ crosses the channel in one of the $\Phi (E)$ transverse modes. The energy dependence of their number (shaded regions) introduces an asymmetry in the channel. The gate potential energetically shifts the modes up ($V_g > 0$) or down ($V_g < 0$), allowing us to tune the response. The picture is valid for weak interactions, as it also applies to Landau quasi-particles.}
    \label{fig:concept}
\end{figure}

\section{Intuitive Landauer Picture}
\label{sec:landauer_picture}

In the absence of interactions, the response can be understood in a Landauer picture [Fig.~\extref[(b)]{fig:concept}]. The total thermoelectric current is a result of two competing effects that favor particle currents in opposite directions.
(i) Since the atom numbers are equal in each reservoir, the chemical potential difference only depends on the difference in temperatures. Increasing temperature lowers the chemical potential as a result of the particle-hole asymmetry of the reservoir density of states around the Fermi energy. This \emph{reservoir asymmetry} prefers particle currents from the cold to the hot side.
(ii) In order to be transferred from one reservoir to the other, a particle of energy $E$ will transit via one of the transverse modes available in the channel, counted by the transport function $\Phi(E)$. The higher the energy, the more modes available, leading to a faster transfer of energetic particles. This \emph{channel asymmetry} favors currents from the hot to the cold reservoir since the hot reservoir has an excess of energetic particles compared to the cold side.

The channel asymmetry is controlled by the gate potential inside the channel as intuitively illustrated in Fig.~\extref[(b)]{fig:concept}: 
For the repulsive case, $V_g > 0$, the channel predominantly allows transport of the most energetic particles present in the hot reservoir. Hence, the net current flows from hot to cold. As the gate potential is made increasingly attractive, particles from the cold reservoir start to contribute and reduce the net current until it vanishes. The precise gate potential where the cancellation happens depends on the initial reservoir conditions and on the transport function $\Phi (E)$. At sufficiently attractive gate potentials, $V_g < 0$, almost all particle energies contribute, and their direction from cold to hot is purely defined by the chemical potential gradient. We expect the intuitive picture to also be valid for weak interactions, as it applies to Landau quasi-particles.

\section{System Dynamics}
\label{sec:dynamics}

Figure~\extref[(a)]{fig:evolution} presents the experimental evolution of the atom number and temperature difference of a non-interacting Fermi gas subject to an initial temperature gradient for different gate strengths. Because of the combined effect of the asymmetries of the reservoirs and the channel, particles flow from the hot to the cold side and build up a negative difference in atom numbers, $\Delta N = N_L - N_R$. Simultaneously, a heat flow drives the temperatures towards equilibrium, thus weakening the thermoelectric current. The particle number difference that builds up induces diffusion that flows against the thermoelectric current, and it vanishes at the turning point. Subsequently, diffusion brings the system to equilibrium. This evolution occurs for all gate potentials, except the most attractive one, $V_g = \SI{-1.06}{\micro\kelvin}$, where the thermoelectric current vanishes. The observed influence of the gate agrees with the intuitive Landauer picture. For more attractive potentials, more transverse modes, counted by the function $\Phi (E)$, open up and permit a faster relaxation. Moreover, the initial response reduces as quantified by the initial particle current $I_N (0)$ in the inset [Appendix~\ref{app_sec:parameter_extraction}]. The current is normalized with the initial bias $\Delta T_0$ to remove variations in the preparation, and it is plotted versus local chemical potential $\mu_\text{loc}$ at the channel center [Appendix~\ref{app_sec:eff_pot}].

The measurements at strong interactions are presented in Fig.~\extref[(b)]{fig:evolution}. As for the non-interacting case, the initial temperature bias is converted into a particle number difference, and it eventually relaxes back to equilibrium. In contrast, the initial and relaxation dynamics with interactions are about three times faster, and the normalized initial currents are enhanced. This result is a consequence of the gate tuning not only the available number of modes but also the density inside the channel. Thus, from repulsive to attractive gate potentials, the gas becomes denser and is expected to eventually cross the superfluid transition in the channel at $\mu_\text{loc} = \SI{0.428}{\micro\kelvin}$ (dashed line in inset); see Appendix~\ref{app_sec:superfluid_transition} and Ref.~\cite{Ku2012}. The combined effect of interactions and attractive gate potential is visible in the blue curve in Fig.~\extref[(b)]{fig:evolution} showing the fast dynamics.

Strikingly, near the critical point, the net initial current reverses its direction from hot to cold (channel dominated) to cold to hot (reservoir dominated), while this effect is absent for an ideal gas in the same parameter regime. In principle, the reversal may also be achieved without interactions as predicted from the Landauer picture and is expected to occur for more attractive gate potentials. However, stronger attractive gates would induce atom losses, which prevent the observation of current reversal.

Contrary to the present work, the current was flowing only in one direction in previous experiments: from hot to cold for a non-interacting two-dimensional gas in the absence of a gate \cite{Brantut2013}, or from cold to hot in the strongly-interacting, quasi-one-dimensional case \cite{Husmann2018}.

As the assumption of a Fermi liquid underlying the Landauer picture breaks down at unitarity, we employ a phenomenological model that captures the transport properties irrespective of the interaction strength. Based on the model, we reason in Sec.~\ref{sec:thermopower} why the reversal occurs with interactions.

\begin{figure*}
    \includegraphics{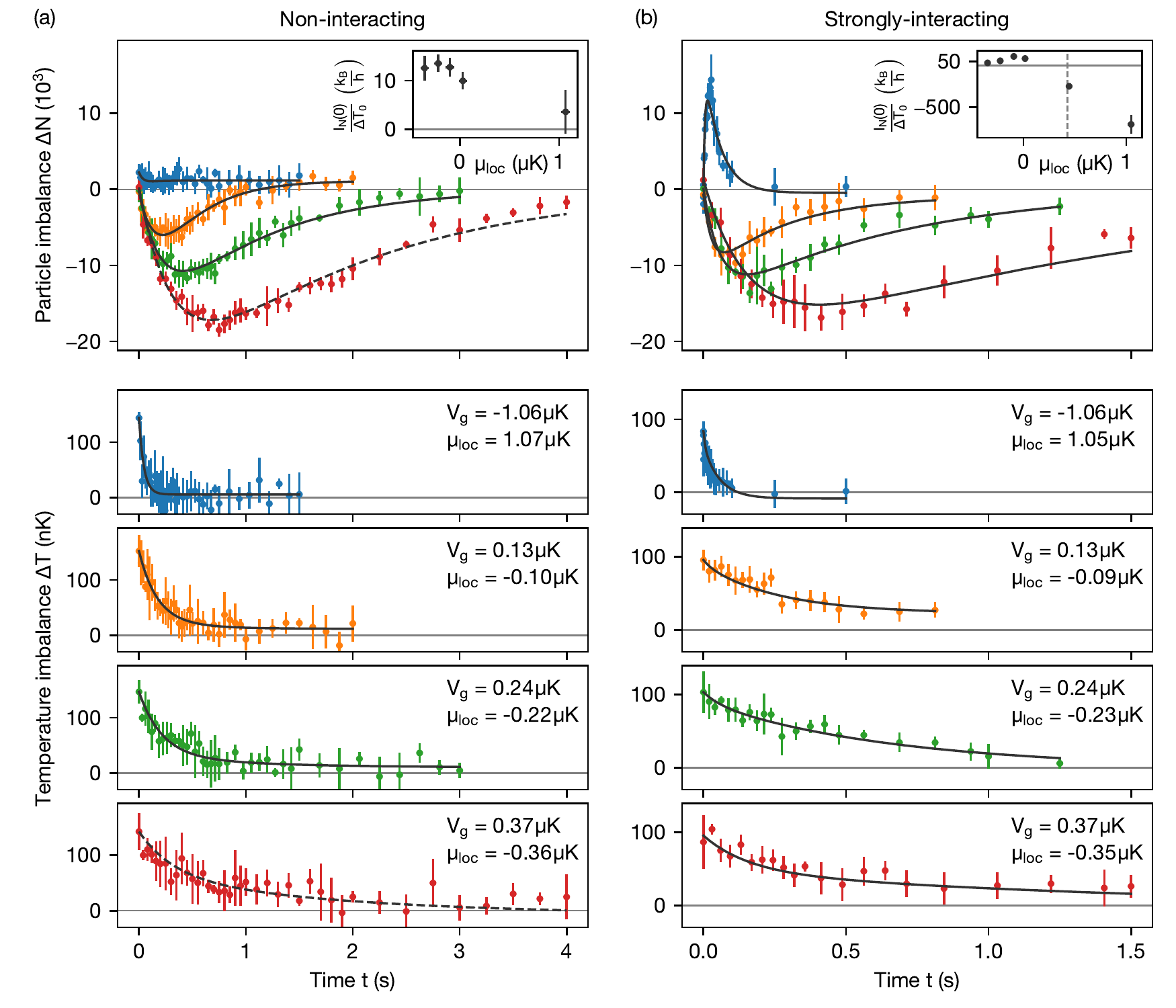}
    \caption{Tuning the thermoelectric response. Temporal evolution of the differences in atom number $\Delta N = N_L - N_R$ and temperature $\Delta T = T_L - T_R$ between the reservoirs of (a) a non- and (b) a strongly-interacting Fermi gas at different gate potentials~$V_g$. Each data point represents a mean over about five repetitions, and error bars show one standard deviation. Lines indicate fits with a phenomenological model involving the transport parameters conductance $G$, thermopower~$\alpha_c$, and Lorenz number~$L$ [Sec.~\ref{sec:phenomenological_model}]. For all estimations, conductance is fixed by a separate measurement (solid lines), except for the dashed line, as it failed to converge otherwise. Insets: Initial particle current~$I_N (0)$ normalized to the prepared temperature difference~$\Delta T_0$ versus local chemical potential~$\mu_\text{loc}$ at the channel center. In the strongly-interacting regime, the vertical dashed line shows the superfluid transition inside the channel while the reservoirs remain uncondensed [Appendix~\ref{app_sec:superfluid_transition}].}
    \label{fig:evolution}
\end{figure*}

\section{Phenomenological Model}
\label{sec:phenomenological_model}

To extract transport properties from the time evolutions, we apply a phenomenological model in linear response. It relates currents of particles $I_N$ and entropy $I_S$ to the differences in chemical potential $\Delta \mu$ and temperature $\Delta T$ \cite{Grenier2016, Goupil2011}, 

\begin{equation}
    \begin{pmatrix}
        I_N \\ I_S
    \end{pmatrix} =
    G
    \begin{pmatrix}
        1 & \alpha_c \\
        \alpha_c & L + \alpha_c^2
    \end{pmatrix}
    \begin{pmatrix}
        \Delta\mu \\ \Delta T
    \end{pmatrix}.
    \label{eq:phenomenological_model}
\end{equation}
As a consequence of the Onsager reciprocal relations, the channel properties are described by only three coefficients: the particle conductance~$G$, the heat conductance~$G_T$, and the Seebeck coefficient or thermopower~$\alpha_c$. The thermopower quantifies the coupling between particle and entropy currents, and the Lorenz number $L = G_T / \bar{T} G$ measures the relative strength of particle and heat conductances.

Thermopower is essential to understand how the competing asymmetries of the reservoirs and the channel appear in the model. At the initial time, the reservoir asymmetry translates a difference in temperature $\Delta T_0$ to a difference in chemical potential $\Delta \mu = - \alpha_r \Delta T_0$ described by the dilatation coefficient $\alpha_r = - ( \partial \mu / \partial T )_{N}$. Together with Eq.~\eqref{eq:phenomenological_model}, the initial response is 
\begin{equation}
	I_N (0) = G \alpha \Delta T_0, 
\end{equation}
where the effective thermopower $\alpha = \alpha_c - \alpha_r$ determines the outcome of the competition between the asymmetries of the channel and the reservoirs, captured by $\alpha_c$ and $\alpha_r$, respectively.

At strong interactions and lowest temperatures, this phenomenological model is expected to fail due to nonlinearities caused by superfluid effects in the reservoirs \cite{Husmann2015}. However, at our elevated temperatures, it captures the behavior well, as observed in Ref.~\cite{Husmann2018}.

The transfer of particles and entropy between the reservoirs via the associated currents $I_N (t) = - \partial_t \Delta N (t) / 2$ and $I_S (t) = - \partial_t \Delta S (t) / 2$ modifies the chemical potential and temperature describing each reservoir and, in turn, the currents. Together with the equation of state in each reservoir, the dynamics are captured by a system of coupled differential equations. Their solution governs how the system evolves from the initial conditions $\Delta N_0$ and $\Delta T_0$ towards equilibrium \cite{Grenier2016} [Appendix~\ref{app_sec:parameter_extraction}], 
\begin{align}
	\Delta N (t) &= \left\{ e_{+} (t) + A e_{-} (t) \right\} \Delta N_0 + B e_{-} (t) \Delta T_0, \\
	\Delta T (t) &= \left\{ e_{+} (t) - A e_{-} (t) \right\} \Delta T_0 + C e_{-} (t) \Delta N_0,
\end{align}
with $e_{\pm} (t) = (\mathrm{e}^{-t / \tau_{+}} \pm \mathrm{e}^{-t / \tau_{-}}) / 2$ and the exponential timescales $\tau_\pm$. Both timescales and the coefficients A, B, and C depend on the transport properties conductance~$G$, thermopower~$\alpha_c$, and Lorenz number~$L$, and on the thermodynamical properties of the reservoirs through compressibility, heat capacity, and dilatation coefficient~$\alpha_r$. The values of the reservoir properties are taken from the measurement [Appendix~\ref{app_sec:extract_thermodynamics}].

We directly extract the transport coefficients from the transients by fitting $\Delta N (t)$ and $\Delta T (t)$ simultaneously, normalized by their statistical uncertainty. To improve accuracy, we reduce the number of free parameters by fixing the conductance to a separately measured value [Appendix~\ref{app_sec:conductance}]. The fits are shown as lines in Fig.~\ref{fig:evolution}. For the dashed curve, conductance was left as a free parameter as it failed to converge otherwise. We discuss the determined transport coefficients in the subsequent chapters.

A conceptually different approach to extract transport coefficients is used in our previous work \cite{Husmann2018}. Instead of utilizing the entire evolutions in $\Delta N (t)$ and $\Delta T (t)$, the method focuses on the turning points in $\Delta N (t)$ where the particle current vanishes. There, for example, the thermopower $\alpha_c = - \Delta \mu / \Delta T$ is the ratio between the differences in chemical potential and temperature [Eq.~\eqref{eq:phenomenological_model}]. Similarly, the heat conductance~$G_T = \bar{T} I_S / \Delta T$ follows by extracting the instantaneous entropy current~$I_S$ via a numerical derivative of $\Delta S (t)$. Therefore, since we measure entire evolutions, it is more natural to fit them directly.

\section{Thermopower}
\label{sec:thermopower}

We first discuss conductance and then present and analyze the results for thermopower. Figures~\extref[(a)]{fig:thermopower} and \extref[(b)]{fig:thermopower} display the measured conductances versus local chemical potential~$\mu_\text{loc}$ in the non- (dark blue) and strongly-interacting (orange) regimes. Conductance is increased by interactions, as previously observed \cite{Stadler2012}, and is a factor of 13(1) larger at the most attractive gate strength [Fig.~\extref[(b)]{fig:thermopower}]. In the non-interacting regime, the solid line represents an \emph{ab initio} prediction that reproduces the measurement well. The model is based on Landauer's theory, which follows the idea that both reservoirs inject particles according to their occupations, and the channel transmits them in one of the available modes \footnote{At the most attractive gate potential, the conductance reaches $75(5)/h$, larger than the number of modes below $\bar{\mu}$ of 40 due to non-zero temperature [Sec.~\ref{sec:setup}].} [Fig.~\extref[(b)]{fig:concept} and Appendix~\ref{app_sec:landauer}]. The light-blue open circle indicates the fitted conductance corresponding to the dashed curve in Fig.~\extref[(a)]{fig:evolution} and is consistently higher than the model and the separate measurement (dark blue point).

The behavior of the thermopower is summarized in Fig.~\extref[(c)]{fig:thermopower}. In the non-interacting regime, the thermopower $\alpha_c$ of the channel (blue points) reduces with increasing local chemical potential towards the dilatation coefficient $\alpha_r$ (horizontal blue line). As expected, the reduction originates from a suppression of the channel asymmetry with more attractive gate strengths, as illustrated in Landauer's picture. Moreover, the dilatation coefficient is a property of the reservoirs and thus unaffected by the gate that locally acts on the channel. The Landauer prediction is indicated as a solid black curve and reproduces well the extracted thermopower in the non-interacting case, except for the most attractive gate strength ($\mu_\text{loc} = \SI{1.1}{\micro\kelvin}$). Theoretically, the reversal of the thermoelectric current is anticipated ($\alpha_c < \alpha_r$), while experimentally it is absent ($\alpha_c \gtrsim \alpha_r$), which might originate from the detailed shape of the confining potential not captured in the theory, such as anharmonicities. Overall, the tunability of the thermopower in the non-interacting regime demonstrates almost full control of the thermoelectric response. In contrast, at unitarity, the thermopower~\(\alpha_c\) becomes smaller than the dilatation coefficient~\(\alpha_r\) for sufficiently attractive gate potentials, and additionally, \(\alpha_r\) and \(\alpha_c\) are reduced compared to the non-interacting case.

The interaction-assisted reduction and reversal can be qualitatively understood from an interpretation based on entropy. On the one hand, the role of the reservoirs is captured by the dilatation coefficient, which can be expressed as the entropy content to add a particle isothermally, $\alpha_r = ( \partial S / \partial N )_{T}$. On the other hand, rewriting Eq.~\eqref{eq:phenomenological_model} as 
\begin{equation}
	I_S = \alpha_c I_N + G_T \frac{\Delta T}{\bar{T}}
\end{equation}
allows for reinterpreting the thermopower of the channel as the average entropy that is reversibly transported by one particle, while the second term captures the irreversible entropy exchange between the reservoirs. At unitarity, pairing correlations reduce the entropy in the spin sector and account for the decrease of $\alpha_c$ and $\alpha_r$ compared to the non-interacting case, as visible in Fig.~\extref[(c)]{fig:thermopower} and in Ref.~\cite{Ku2012}. Because the attractive gate increases the density and therefore the interaction effects at the center, we expect further reduction of the thermopower~$\alpha_c$.

This argument suggests a way to estimate where the current reverses in the presence of interactions. Since the interparticle collision rate is enhanced at unitarity, we assume the gas to be locally in equilibrium at the center [Appendix~\ref{app_sec:boltzmann}]. This allows us to think of the thermopower $\alpha_c$ as the dilatation coefficient inside the channel at temperature $\bar{T}$ and chemical potential $\mu_\text{loc}$ leading to a reversal expected at $\mu_\text{loc} \sim \SI{-0.1}{\micro\kelvin}$. The value is relatively close to the observed location despite the simplicity of the estimation and the neglect of the transverse mode structure.

\begin{figure}
    \includegraphics{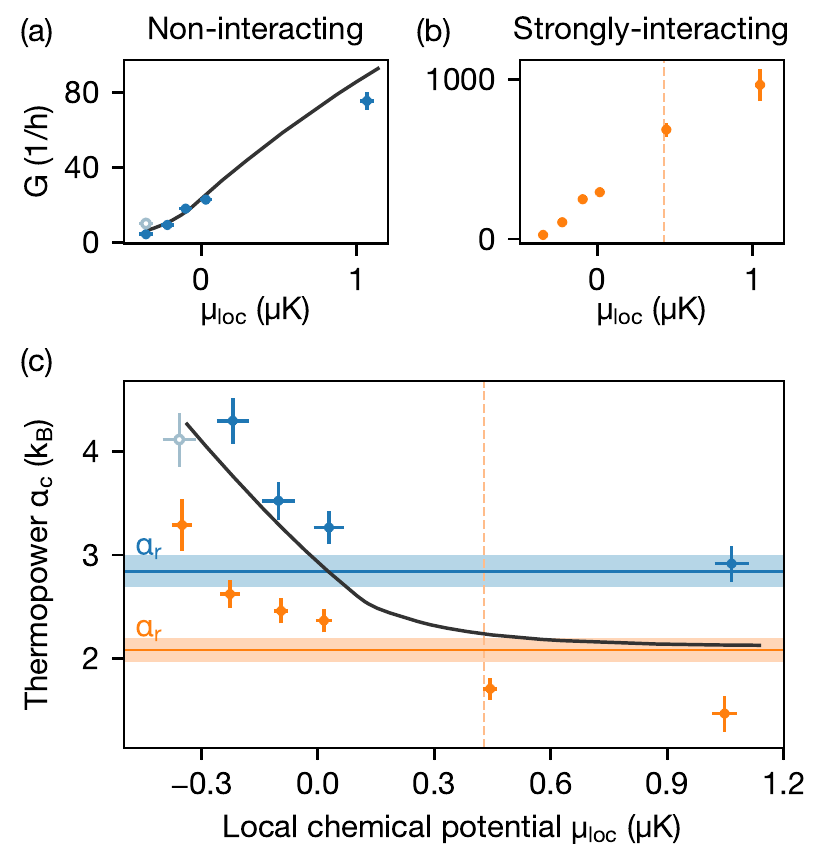}
    \caption{Controlling thermopower with a gate potential. (a,b) Conductance~$G$ versus local chemical potential~$\mu_\text{loc}$ in the (a) non- and (b) strongly-interacting regimes. Separately measured values are indicated as dark points [Appendix~\ref{app_sec:conductance}], and the light-blue open circle shows the fitted conductance corresponding to the dashed curve in Fig.~\extref[(a)]{fig:evolution}. (c) Fitted thermopower~$\alpha_c$ without (blue points) and with strong (orange points) interparticle interactions and the corresponding dilatation coefficients~$\alpha_r$ of the reservoirs (horizontal lines). Shaded regions indicate one standard deviation to each side. Solid black lines show an \emph{ab initio} prediction based on Landauer theory in the absence of interactions [Appendix~\ref{app_sec:landauer}]. The vertical dashed line locates the superfluid transition inside the channel, while the reservoirs remain uncondensed [Appendix~\ref{app_sec:superfluid_transition}]. No measurement was taken at $\mu_\text{loc} \sim \SI{0.4}{\micro\kelvin}$ in the non-interacting case.}
    \label{fig:thermopower}
\end{figure}

\section{Lorenz Number}
\label{sec:lorenz_number}

The Lorenz number $L = G_T / \bar{T} G$ compares the ability of systems to conduct heat and particles and indicates whether Fermi liquid behavior is present. In this case, transport is described by Landau quasi-particles that carry both charge and energy, leading to a constant Lorenz number, $L_\text{WF} = \pi^2 / 3 \cdot k_B^2$, which is known as the Wiedemann-Franz law.

\begin{figure}
    \includegraphics{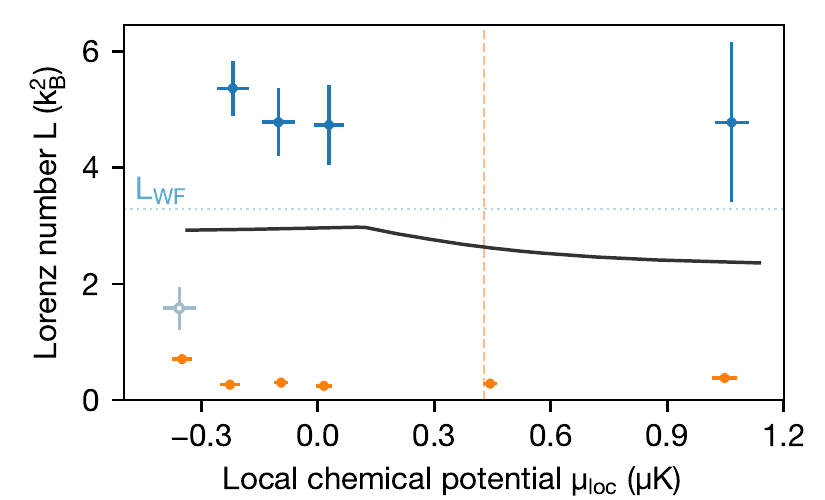}
    \caption{Lorenz number~$L$ versus local chemical potential~$\mu_\text{loc}$ at the channel center. Fitted values in the non- and strongly-interacting regimes are shown as blue and orange points, respectively. The theoretical prediction based on Landauer theory is indicated by a solid black curve [Appendix~\ref{app_sec:landauer}], and the Wiedemann-Franz value $L_\text{WF} = \pi^2 / 3 \cdot k_B^2$ valid for degenerate Fermi liquids is shown by a blue horizontal line. In the strongly-interacting case, the vertical dashed line locates the superfluid transition inside the channel, while the reservoirs remain uncondensed [Appendix~\ref{app_sec:superfluid_transition}].}
    \label{fig:lorenz}
\end{figure}

Figure~\ref{fig:lorenz} displays the fitted Lorenz number versus local chemical potential in the non- (blue dots) and strongly-interacting (orange dots) cases. Without interactions, the extracted values are consistently higher than $L_\text{WF}$ and also higher than the Landauer theory (solid line). The theory considers our mesoscopic geometry at finite temperature and approaches $L_\text{WF}$ with increasing degeneracy. Deviations from the Wiedemann-Franz law are found in other systems with either increased \cite{Wakeham2011} or decreased \cite{Lee2017, Bruin2013, Hartnoll2015} numbers. Here, we partly attribute the inconsistency to small systematic shifts in the measured temperature difference, which mostly affect the Lorenz number [Appendix~\ref{app_sec:conductance}].

Despite the challenging absolute estimation, the Lorenz number is reduced by one order of magnitude when increasing the interactions to unitarity. As the fitted heat conductance is relatively insensitive to the interaction strength, this decrease can be mostly attributed to the enhancement of the conductance seen in Figs.~\extref[(a)]{fig:thermopower} and \extref[(b)]{fig:thermopower}. Also, the smaller uncertainties on the Lorenz number stem from the enhanced conductances and give us confidence in thinking that the Wiedemann-Franz law is violated here, as experimentally observed in a one-dimensional geometry \cite{Husmann2018} and theoretically supported in Refs.~\cite{Pershoguba2019_0, Uchino2020}.

\section{Discussion}
\label{sec:discussion}

Mesoscopic transport properties are influenced by the geometry of the structure \cite{Imry2002}; thus, it is instructive to compare our findings at unitarity with a one-dimensional channel measured in similar conditions \cite{Husmann2018}. There, the narrow geometry blocked irreversible heat currents, leading to a non-equilibrium steady state with a finite temperature difference. Relaxation was restored by widening the channel, thanks to an enhanced heat exchange. This finding agrees with our results for a wide quasi-two-dimensional channel, where we observe relaxation to equilibrium and a heat conductance that is similar to the non-interacting case. In both works, the Lorenz number is reduced with interactions by an order of magnitude. However, the reasons are different: In Ref.~\cite{Husmann2018}, it stems from a reduced heat conductance, while here it is almost purely an effect of enhanced particle conductance. Thermopower is reduced by interactions in our wide, quasi-two-dimensional geometry as qualitatively explained by pairing correlations that restrict average entropy transfer per particle. In contrast, in the quasi-one-dimensional channel, it shows the surprising behavior of following the non-interacting prediction, which so far has eluded any explanation \cite{Pershoguba2019_0}.

In summary, we control the magnitude and direction of thermoelectric currents through a mesoscopic structure in the presence of weak and strong interparticle interactions. In our parameter regime at weak interactions, particles are flowing consistently in one direction, while at unitarity, we observe a striking interaction-assisted reversal. We explain the reversal by a competition of reservoir and channel parameters, which are affected by pairing correlations that are expected in the large critical region around the superfluid transition \cite{Debelhoir2016}. Indeed, the reversal occurs before the normal-to-superfluid transition, and its precise location depends on the geometry of the system and the reservoir conditions via thermopower and the dilatation coefficient. Unitary Fermi gases are not the only system where thermopower can be affected by interactions: In a two-dimensional Bose gas, the Seebeck coefficient changes its sign close to the superfluid transition \cite{Hazlett2013}, and in two-dimensional electron gases, interactions are predicted to reduce the thermopower and potentially reverse its sign \cite{Dolgopolov2011}.

The option to induce thermoelectric currents in either direction is appealing when considering our dynamics as an open thermodynamic cycle. In our system, an atomic flow from hot to cold acts against the chemical potential bias and converts heat into work as a thermoelectric engine. Inversely, the system acts as a thermoelectric cooler when the flow transfers heat from cold to hot. The initial direction of the current therefore determines which mode of operation takes place before the other: a thermoelectric engine in the channel-dominated regime or a cooler in the reservoir-dominated regime. In both modes, the conversion efficiency of the cycles is characterized by the figure of merit $ZT = \alpha_c^2 / L$ \cite{Goupil2011}. Contrary to a non-interacting system, interactions strongly reduce the Lorenz number while thermopower remains at a similar order of magnitude. Overall, we estimate that interactions improve the figure of merit by a factor of 7(3), showing the relevance of strongly correlated quantum materials for thermoelectric applications.

Our system readily allows us to probe the thermoelectric response of more complicated structures. Drawing from the technique to imprint local effective Zeeman shifts \cite{Lebrat2019}, it opens new perspectives on coupling spin and heat transport \cite{Bauer2012}. By adding strong correlations, intriguing thermoelectric effects could be observed \cite{Ozaeta2014, Bergeret2018}.

\begin{acknowledgments}

We thank Thierry Giamarchi, Leonid Glazman, René Monnier, Shun Uchino, and Anne-Maria Visuri for fruitful discussions; Alexander Frank for electronic support; and Mohsen Talebi for a careful reading of the manuscript. We thank Daniel Kestner for support and the source code regarding the rank order fitting method. We acknowledge the Swiss National Science Foundation (Grants No. 182650 and No. NCCR-QSIT) and European Research Council advanced grant TransQ (Grant No. 742579) for funding. L.C. is supported by an ETH Zurich Postdoctoral Fellowship, the Marie Curie Actions for People COFUND program, and the European Union Horizon 2020 Marie Curie TopSpiD program (Grant No. 746150).
\end{acknowledgments}

\appendix

\section{Experimental Details}
\label{app_sec:experimental_details}

A cloud of fermionic lithium atoms is prepared in a balanced mixture of the two lowest hyperfine states in a hybrid trap. In the transverse directions ($x$, $z$), the trap is formed by a far-detuned laser beam at a wavelength of \SI{1064}{\nano\meter} and longitudinally ($y$) by a quadratic magnetic field. We evaporatively cool the elongated cloud by reducing the transverse confinement either on a broad Feshbach resonance at \SI{832}{G} when preparing a unitary gas or at \SI{431}{G} when preparing a non-interacting gas. Subsequently, the magnetic field is kept on resonance or ramped to \SI{568}{G} where the scattering length reaches $173 \, a_0$ as expressed with Bohr's radius $a_0$. The small and non-zero scattering length ensures that the reservoirs are thermalized in the nearly non-interacting regime. Moreover, the transport properties at these weak interactions are expected to be equal to the non-interacting case, as previously found in Ref.~\cite{Krinner2015}. Finally, the optical trap is recompressed, leading to the confinement frequencies during transport $\nu_{tx} = \SI{232(1)}{\hertz}$ and $\nu_{tz} = \SI{212(1)}{\hertz}$ in the transverse direction and longitudinally $\nu_{ty}$ of \SI{26.6(2)}{\hertz} in the non-interacting case and \SI{32.2(2)}{\hertz} at unitarity.

The transport channel is imprinted by a repulsive laser beam at $\SI{532}{\nano\meter}$, which is shaped by a 0-$\pi$ phase plate into a TEM$_{01}$ mode and focused onto the center of the cloud. This compresses the cloud in the vertical $z$ direction, resulting in a harmonic confinement of $\nu_{cz} =\SI{4.5(8)}{\kilo\hertz}$. Its $1/e^2$ waists are, in the longitudinal direction, $w_{cy} =\SI{30.2}{\micro\meter}$ and, transversally, $w_{cz} = \SI{9.5}{\micro\meter}$.

We prepare the initial conditions of the two reservoirs in two steps. First, the cloud is centered on the transport channel using a magnetic gradient to make the atom numbers on both sides equal; then, we split the reservoirs with an elliptical repulsive laser beam. Second, a temperature difference is created by heating one side with a red-detuned beam at \SI{767}{\nano\meter} focused into one reservoir and modulating its intensity parametrically. The experimentally optimized modulation frequency is \SI{547}{\hertz}, which is on the order of the transverse confinement frequencies $\nu_{tx}$ and $\nu_{tz}$. The beam position is controlled with a piezo-mirror and can be directed into either reservoir. After letting the reservoirs thermalize for \SI{10}{\milli\second} we reach the initial conditions reported in Table~\ref{app_tab:exp_conditions}. 

The attractive gate potential is created with the same beam used for heating and requires a careful alignment onto the channel center. Its waists are $w_{gx} = \SI{34.3(2)}{\micro\meter}$ and $w_{gy} = \SI{33.5(2)}{\micro\meter}$. The repulsive one is formed by an elliptical beam at $\SI{532}{\nano\meter}$ that is focused onto our channel with waist $w_{gx} = \SI{53.66(3)}{\micro\meter}$ in the transversal direction, and along the transport $w_{gy} = \SI{8.58(3)}{\micro\meter}$. Besides acting as a gate, the same beam enables and blocks transport at large powers in a controlled way. After the time~$t$, the reservoirs are again separated, and an absorption picture is taken after a short time of flight of $\SI{1}{\milli\second}$. This reduces the central densities and allows us to image with intensities well below saturation. From the absorption image, we extract the thermodynamical properties as described in Appendix~\ref{app_sec:extract_thermodynamics}.

\begin{table}
	\caption{Reservoir conditions for measuring thermoelectricity in the non- and strongly-interacting regimes. The values are averaged over the realizations at different gate potentials, and their uncertainties indicate one standard deviation.} \vspace{10pt}
	\label{app_tab:exp_conditions}
	\begin{tabular}{lSS}
		\textbf{Quantity} & \textbf{Non-interacting} & \textbf{Unitary} \\
		\hline
		$N$ ($10^3$) & 121(2) & 107(4) \\
		\hline
		$T_L$ (\si{\nano\kelvin}) & 281(8) & 220(5) \\
		$T_R$ (\si{\nano\kelvin}) & 134(7) & 129(3) \\
		\hline
		$\mu_L$ ($k_B \, \si{\nano\kelvin}$) & -43(27) & 65(13) \\
		$\mu_R$ ($k_B \, \si{\nano\kelvin}$) & 344(12) & 238(8) \\
		\hline
		$\bar{T}$ (\si{\nano\kelvin}) & 208(6) & 174(2) \\
		$\bar{\mu}$ ($k_B \, \si{\nano\kelvin}$) & 151(16) & 152(9) \\
		\hline
		$\Delta T$ (\si{\nano\kelvin}) & 147(11) & 90(7) \\
		$\Delta \mu$ ($k_B \, \si{\nano\kelvin}$) & -387(29) & -172(13) \\
		\hline
	\end{tabular}
\end{table}

\section{Data Analysis}

\subsection{Thermodynamics}
\label{app_sec:thermodynamics}

Throughout this Appendix, the equations of state of a non-interacting and unitary Fermi gas are used in the homogeneous and trapped cases. This section summarizes the relevant thermodynamical relations.

\subsubsection{Homogeneous Unitary Fermi Gas}
\label{app_sec:homogeneous_unitary_gas}

The universality hypothesis states that the homogeneous unitary Fermi gas is described by the interatomic distance and the thermal wavelength only \cite{Ho2004}. This restricts the equation of state to the form \cite{Hou2013}
\begin{equation}
	n \lambda_T^3 = f_n (q), 
\end{equation}
with the particle density $n$, the thermal wavelength $\lambda_T = \sqrt{2 \pi \hbar^2 / m k_B T}$, and the dimensionless function $f_n$ depending on the reduced chemical potential $q = \beta \mu$. The scaling function $f_n$ was measured around the superfluid transition \cite{Ku2012} and theoretically extended in the degenerate and thermal limits with a phonon model and a third order virial expansion, respectively \cite{Hou2013}\footnote{The phonon model [Eq.~(37) in Ref.~\cite{Hou2013}] and the third-order virial expansion [Eq.~(44) in Ref.~\cite{Hou2013}] include both hyperfine states, while our quantities refer to one of them. In the intermediate regime, the scaling function is obtain from Fig.~4A in Ref.~\cite{Ku2012} using the equation of state at unitarity $n \lambda_T^3 = f_n (q)$ and without interactions $n_0 \lambda_T^3 = F_{1/2} (q)$. The function $F_j$ denotes the complete Fermi-Dirac integral of order $j$ [Eq.~(25.12.14) in Ref.~\cite{Olver2010}].}. Using the equation of state, the normalized temperature is given as 
\begin{equation}
	\frac{T}{T_F} = \left( \frac{4}{3 \sqrt{\pi} f_n (q)} \right)^{2/3}, 
	\label{app_eq:homogeneous_unitary_gas_T_over_TF}
\end{equation}
with the Fermi energy $E_F = k_B T_F = \hbar^2 / (2 m) (6 \pi^2 n)^{2/3}$ \cite{Hou2013}. This result links the normalized temperature $T / T_F$ and the chemical potential $\mu / E_F = T / T_F \cdot q$ to each other. Below the critical temperature $T_c = 0.167 \, T_F$, the gas becomes superfluid as observed in Ref.~\cite{Ku2012}.

\subsubsection{Harmonically Trapped Fermi Gas}
\label{app_sec:harmonic_fermi_gas}

In a harmonic trap, the equilibrium properties are captured by the geometric mean $\bar{\nu} = (\nu_x \nu_y \nu_z)^{1/3}$ of the confinement frequencies, the Fermi energy $E_F = h \bar{\nu} (6 N)^{1/3}$, and the reduced chemical potential $q_0 = \beta \mu_0$ at the trap center. For our work, the total atom number $N$ in each hyperfine state, the compressibility $\kappa = (\partial N / \partial \mu)_T$, the dilatation coefficient $\alpha_r = - ( \partial \mu / \partial T )_{N}$, the specific heat $C_N = T (\partial S / \partial T)_N$ at fixed atom number, and the internal energy $U$ are relevant. Following Ref.~\cite{Lebrat2019_0}, they are given in the non-interacting case by
\begin{align}
	N &= \left( \frac{k_B T}{h \bar{\nu}} \right)^3 F_2 (q_0), \\
	\frac{\kappa}{N} &= \frac{1}{k_B T} \frac{F_1 (q_0)}{F_2 (q_0)}, \\
	\frac{\alpha_r}{k_B} &= 3 \frac{F_2 (q_0)}{F_1 (q_0)} - q_0, \\
	\frac{C_N}{N k_B} &= 12 \frac{F_3 (q_0)}{F_2 (q_0)} - 9 \frac{F_2 (q_0)}{F_1 (q_0)}, \\
	\frac{U}{N E_F} &= \frac{3}{6^{1/3}} \frac{F_3 (q_0)}{F_2 (q_0)^{4/3}}, \label{app_eq:harmonic_nif_U} \\
	\frac{T}{T_F} &= ( 6 F_2(q_0))^{-1/3},
\end{align}
with $F_j$ symbolizing the complete Fermi-Dirac integral of order $j$ [see Eq.~(25.12.14) in Ref.~\cite{Olver2010}]. At unitarity, the relations are stated in Ref.~\cite{Husmann2018_0}, and they read as
\begin{align}
	N &= \frac{4}{\sqrt{\pi}} \left( \frac{k_B T}{h \bar{\nu}} \right)^3 N_2 (q_0), \\
	\frac{\kappa}{N} &= \frac{1}{2 k_B T} \frac{N_0 (q_0)}{N_2 (q_0)}, \\
	\frac{\alpha_r}{k_B} &= 6 \frac{N_2 (q_0)}{N_0 (q_0)} - q_0, \\
	\frac{C_N}{N k_B} &= 8 \frac{N_4 (q_0)}{N_2 (q_0)} - 18 \frac{N_2 (q_0)}{N_0 (q_0)}, \\
	\frac{U}{N E_F} &= \left( \frac{\pi}{9} \right)^{1/6} \frac{N_4 (q_0)}{N_2 (q_0)^{4/3}}, \label{app_eq:harmonic_unitary_U} \\
	\frac{T}{T_F} &= \left( \frac{24}{\sqrt{\pi}} N_2(q_0) \right)^{-1/3},
\end{align}
with the integral $N_j (q) = \int_0^\infty \! r^j f_n (q - r^2) \,\mathrm{d}r$ involving the scaling function $f_n$ defined in Appendix~\ref{app_sec:homogeneous_unitary_gas}.

To describe the properties of a single reservoir, we need to divide the extensive quantities for the full trap by two. 
This division applies to the extensive variables atom number~$N$, compressibility~$\kappa$, specific heat~$C_N$, and the internal energy~$U$.

\subsubsection{Extracting Thermodynamical Properties}
\label{app_sec:extract_thermodynamics}

The basis to extract thermodynamical properties of each reservoir $r$ from the absorption images is the virial theorem \cite{Guajardo2013}. It holds for non-interacting and unitary \cite{Thomas2005} Fermi gases and relates the internal energy per particle $U_r / N_r = 3 m \omega_{ty}^2 \langle y^2 \rangle$ to the second moment $\langle y^2 \rangle = \int \! y^2 n_r (y) \,\mathrm{d}y / N_r$ in the transport direction. The one-dimensional density~$n_r (y)$ can be directly deduced from the absorption images by summing over the transverse direction. Together with the total atom number~$N_r$, we obtain the Fermi energy $E_{F,r}$ and, subsequently with Eq.~\eqref{app_eq:harmonic_nif_U} or \eqref{app_eq:harmonic_unitary_U}, the reduced chemical potential $\beta \mu_0$ at the trap center. Then, the temperature $T_r$ and all other thermodynamical properties of the reservoirs can be derived from the formulas in Appendix~\ref{app_sec:harmonic_fermi_gas}.

\subsection{Parameter Extraction}
\label{app_sec:parameter_extraction}

First, we explain the linear phenomenological model \cite{Grenier2016} and, second, its application to extract the transport parameters from the measured evolutions [Fig.~\ref{fig:evolution}].

\subsubsection{Phenomenological Model}

The evolutions in the differences in atom number and temperature are phenomenologically modeled by the response matrices of the channel and the reservoirs. The channel reacts to the applied biases with currents of particles and entropy [Eq.~\eqref{eq:phenomenological_model}] described by 
\begin{equation}
    \begin{pmatrix}
        I_N \\ I_S
    \end{pmatrix} =
    G
    \begin{pmatrix}
        1 & \alpha_c \\
        \alpha_c & L + \alpha_c^2
    \end{pmatrix}
    \begin{pmatrix}
        \Delta\mu \\ \Delta T
    \end{pmatrix}, 
    \label{app_eq:force_flux}
\end{equation}
with the transport coefficients conductance~$G$, thermopower~$\alpha_c$, and Lorenz number~$L$. The reservoir thermodynamics are described in linear response by 
\begin{equation}
    \begin{pmatrix}
        \Delta N \\ \Delta S
    \end{pmatrix} =
    \kappa 
    \begin{pmatrix}
        1 & \alpha_r \\
        \alpha_r & l + \alpha_r^2
    \end{pmatrix}
    \begin{pmatrix}
        \Delta\mu \\ \Delta T
    \end{pmatrix}.
    \label{app_eq:reservoirs}
\end{equation}
The involved thermodynamical quantities of each reservoir are the compressibility~$\kappa$, the dilatation coefficient~$\alpha_r$, and the reservoir analogue of the Lorenz number $l = C_N / \bar{T} \kappa$, which depends on the heat capacity $C_N$. Within linear response, they are constant throughout the evolution and are evaluated at the equilibrium chemical potential~$\bar{\mu}$ and temperature $\bar{T}$ [Appendix~\ref{app_sec:harmonic_fermi_gas}]. The two sets of equations are related via $I_N (t) = - \partial_t \Delta N (t) / 2$ and $I_S (t) = - \partial_t \Delta S (t) / 2$, which show that changes in the reservoirs originate from currents leaving and entering them. Altogether, they form a system of two coupled linear differential equations, with the solution given by
\begin{align}
	\Delta N (t) &= \left\{ e_{+} (t) + A e_{-} (t) \right\} \Delta N_0 + B e_{-} (t) \Delta T_0, \\
	\Delta T (t) &= \left\{ e_{+} (t) - A e_{-} (t) \right\} \Delta T_0 + C e_{-} (t) \Delta N_0, 
\end{align}
with $e_{\pm} (t) = (\mathrm{e}^{-t / \tau_{+}} \pm \mathrm{e}^{-t / \tau_{-}}) / 2$ and the timescales $\tau_\pm = \tau_0 / \lambda_\pm$ and $\tau_0 = \kappa / 2 G$. The eigenvalues of the evolution matrix describing the system of differential equations are 
\begin{equation}
	\lambda_\pm = \frac{1}{2} \left( 1 + \frac{L + \alpha^2}{l} \right) \pm \sqrt{\frac{1}{4} \left(1 + \frac{L + \alpha^2}{l} \right)^2 - \frac{L}{l}}.
\end{equation}

The coefficients A, B, and C depend on the transport and reservoir properties 
\begin{align}
	A &= \frac{1 - (L + \alpha^2) / l}{\lambda_{+} - \lambda_{-}}, \\
	B &= \frac{2 \kappa \alpha}{\lambda_{+} - \lambda_{-}}, \\
	C &= \frac{2 \alpha}{\kappa l (\lambda_{+} - \lambda_{-})}, 
\end{align}
with the effective thermopower $\alpha = \alpha_c - \alpha_r$. Note that these equations correct the typographical errors present in Ref.~\cite{Grenier2016}.

\subsubsection{Data Fitting}

We extract the transport parameters by simultaneously fitting the measured evolutions to the phenomenological model using the standard least-squares method. It minimizes the following sum of the squared residuals: 
\begin{equation}
	\chi^2 = \sum_i \left[ \left( \frac{\Delta \tilde{N} (t_i) - \Delta \tilde{N}_i}{\sigma_{N}} \right)^2 + 
	\left( \frac{\Delta \tilde{T} (t_i) - \Delta \tilde{T}_i}{\sigma_{T}} \right)^2 \right].
\end{equation}
The simultaneous fitting requires normalizing the residuals with the respective uncertainties $\sigma_N$ and $\sigma_T$, which we determine as an average over the standard deviations at the individual times. The measured differences $\Delta \tilde{N}_i$ and $\Delta \tilde{T}_i$ at time~$t_i$ include small offsets due to imperfections in the calibrations. We account for these by shifting the model evolutions $\Delta N (t)$ and $\Delta T (t)$ by constants and obtain $\Delta \tilde{N} (t) = \Delta N (t) + \Delta N_\text{off}$ and $\Delta \tilde{T} (t) = \Delta T (t) + \Delta T_\text{off}$, respectively.

In the model, the reservoir properties $\kappa$, $\alpha_r$, and $l$ are set to the values directly extracted from the absorption images [Appendix~\ref{app_sec:extract_thermodynamics}]. Moreover, the conductance $G$ is fixed by a separate measurement [Appendix~\ref{app_sec:conductance}], which reduces the number of free parameters and improves the fit. 

We determine $\Delta N_0$ and $\Delta T_0$ from the corresponding measured initial values. The offset $\Delta N_\text{off}$ either follows from an average over points where the evolution is relaxed, or it is taken to be the same as the initial value. In contrast, the offset $\Delta T_\text{off}$ cannot be easily determined for evolutions that do not completely relax within our measurement. Thus, we fit it consistently for all dynamics and checked, with evolutions where the temperature completely relaxes, that the results do not depend on whether the offset $\Delta T_\text{off}$ is fixed or not. In summary, the parameters $\alpha$, $L$, and $\Delta T_\text{off}$ are free in the model. Exceptionally, for the non-interacting curve at $V_g = \SI{0.37}{\micro\kelvin}$, the conductance is also free, as otherwise the fit fails to converge [Fig.~\ref{fig:evolution}].

To analyze the parameter estimation, we exemplify the sum of the squared residuals $\chi^2$ versus thermopower~$\alpha$ and Lorenz number $L$ for the gate potential $V_g = \SI{0.13}{\micro\kelvin}$ [Figs.~\extref[(a)]{app_fig:error_landscape} and \extref[(b)]{app_fig:error_landscape}]. For both interaction strengths, there exists a single, isolated global minimum that is found by the Levenberg-Marquardt algorithm (gray point). Its uncertainty only includes the standard deviations $\sigma_N$ and $\sigma_T$ and assumes the fixed parameters to be precisely known. However, their uncertainties will lead to a change in the optimal values for $\alpha$ and $L$. Figures~\extref[(c)]{app_fig:error_landscape} and \extref[(d)]{app_fig:error_landscape} show a plot of these optimal values when one fixed parameter is varied within its uncertainty. These variations are mostly symmetrical and thus leave the values unaffected but increase the overall error bar in thermopower and Lorenz number.

\begin{figure*}
    \includegraphics{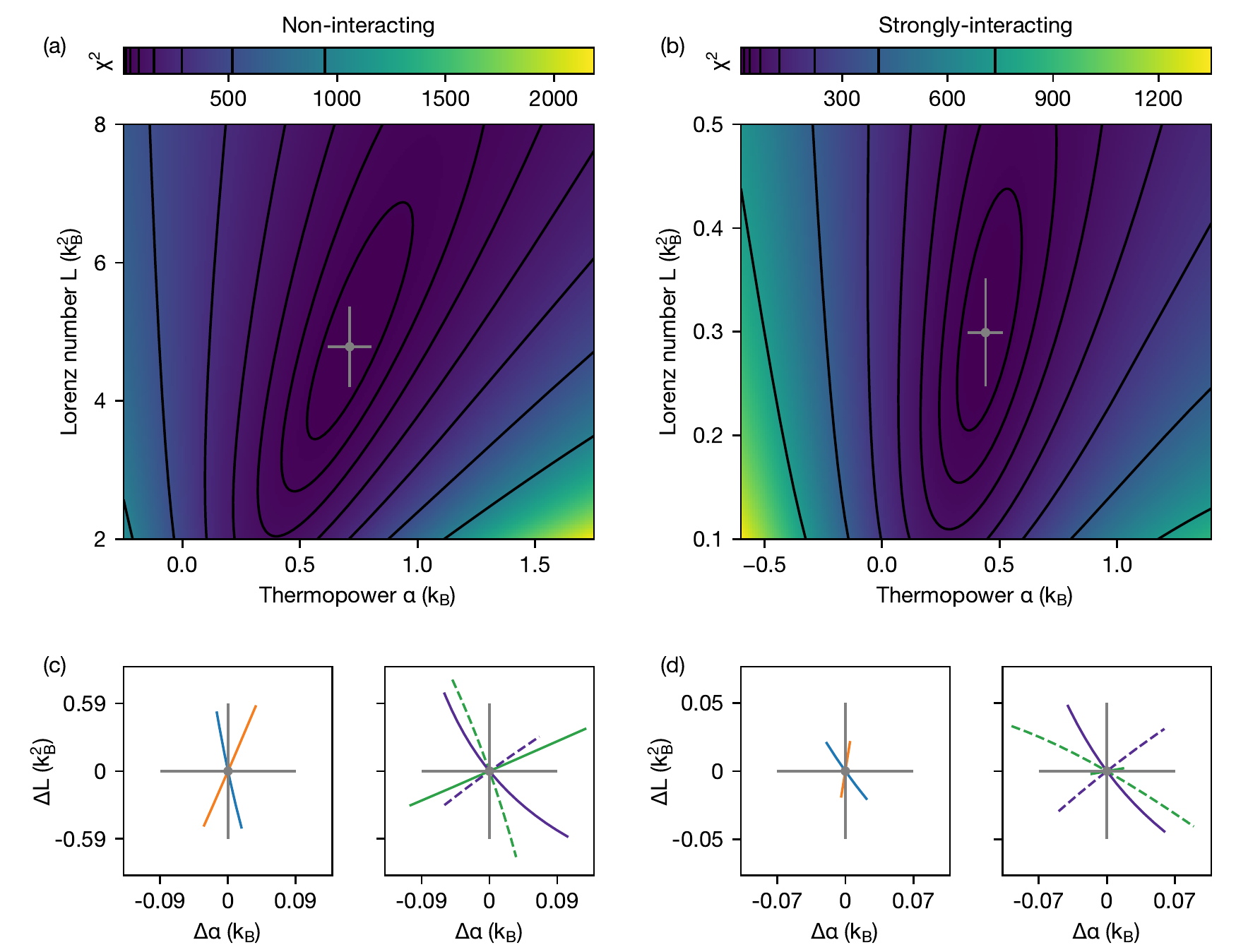}
    \caption{Transport parameter estimation. (a,b) Sum of squared residuals~$\chi^2$ versus effective thermopower~$\alpha$ and Lorenz number~$L$ in the (a) non- and (b) strongly-interacting regimes at $V_g = \SI{0.13}{\micro\kelvin}$. Contour lines highlight the behavior at logarithmically spaced level values indicated by vertical black lines in the color bar. The offset~$\Delta T_\text{off}$ in the temperature difference is fixed to the fitted value. The optimal fit parameters and uncertainties are indicated by the point and the error bar, respectively.
    (c,d) Variations of the optimal thermopower and Lorenz number when one fixed fit parameter is changed within its uncertainty (colored lines). Left panel: modifications in conductance (orange) and compressibility (blue). Right panel: modifications in initial values $\Delta N_0$ (solid green), $\Delta T_0$ (solid violet), and the offsets $\Delta N_\text{off}$ (dashed green) and $\Delta T_\text{off}$ (dashed violet). Changes in the thermodynamic analogue of the Lorenz number $l$ are too small to be visible.}
    \label{app_fig:error_landscape}
\end{figure*}

It is known that estimating parameters of exponential models is challenging, as small changes in the data can strongly influence the parameters \cite{OLeary2013}. Thus, in addition to the least-squares method, we used a technique based on rank order that is robust against outliers and shown to improve the estimation for exponential models \cite{Ierley2019, Kestner2020}. Applying this method led to comparable results; hence, we conclude that, for our model and data, the least-squares method performs well. As the estimation with the rank order fit is numerically demanding due to many local, closely spaced minima \cite{Ierley2019}, we report the values with the least-squares method, which gives a single, well-defined minimum.

Based on the fitted results, the initial response $I_N (0) / \Delta T_0 = G \alpha$ is shown in the insets of Fig.~\ref{fig:evolution}, the thermopower of the channel $\alpha_c = \alpha + \alpha_r$ in Fig.~\ref{fig:thermopower}, and the Lorenz number $L$ in Fig.~\ref{fig:lorenz}.

\subsection{Conductance Measurement}
\label{app_sec:conductance}

To measure conductances, we prepare the reservoirs at different particle numbers and equal temperatures and study their exponential decay towards equilibrium. We fit the characteristic time~$\tau_0$ and, together with the compressibility $\kappa$ of each reservoir [see details in Appendix~\ref{app_sec:thermodynamics}], conductance follows from the relation $\tau_0 = \kappa / 2 G$ analogous to a discharging capacitor \cite{Krinner2015}. The conductances are measured at the same average chemical potential~$\bar{\mu}$ and temperature~$\bar{T}$ as the thermoelectric responses to allow comparison between them. The conductances are shown in Figs.~\extref[(a)]{fig:thermopower} and \extref[(b)]{fig:thermopower}.

The conductances are extracted under the assumption that both reservoirs are at equal temperatures. However, we suspect that in our measured temperature differences, an offset is present, as visible in Fig.~\extref{fig:evolution} when the system is relaxed. During preparation, we heat both reservoirs up to seemingly the same temperature~$\bar{T}$, while the measurement offset leads to a physical temperature difference. In turn, it induces a thermoelectric current with the value 
\begin{equation}
	I_N = G \, \frac{\Delta N}{\kappa} 
	\left( 1 + \kappa \alpha \frac{\Delta T}{\Delta N} \right), 
\end{equation}
which follows from Eqs.~\eqref{app_eq:force_flux} and \eqref{app_eq:reservoirs} and the thermopower $\alpha = \alpha_c - \alpha_r$. The second summand in the brackets quantifies the relative deviations, which are estimated at around \SI{10}{\percent} for the non-interacting case at a gate potential $V_g$ of \SI{0.13}{\micro\kelvin}. We use the compressibility $\kappa = \SI{22}{\second\per h}$, the thermopower $\alpha = 0.7 k_B$, the particle number difference $\Delta N = \num{48e3}$, and the residual temperature bias $\Delta T \sim - \SI{15}{\nano\kelvin}$.

A systematically increased conductance mostly reduces the fitted Lorenz number and only slightly decreases the thermopower, as is visible in Figs.~\extref[(c)]{app_fig:error_landscape} and \extref[(d)]{app_fig:error_landscape}. The Lorenz number is reduced to around $4.2 k_B^2$ for a gate potential of $V_g = \SI{0.13}{\micro\kelvin}$ and similarly for the other non-interacting conditions. Thus, the systematic shifts in the conductance can partly explain the deviations from the Wiedemann-Franz value $L_\text{WF} = \pi^2 / 3 \cdot k_B^2$ in the non-interacting measurements.

\subsection{Superfluid Transition at Unitarity}
\label{app_sec:superfluid_transition}

Superfluidity inside the reservoirs or the channel can strongly influence the transport properties, and thus, it is important to characterize the gas in these locations in the strongly-interacting regime.

\begin{figure}
    \includegraphics{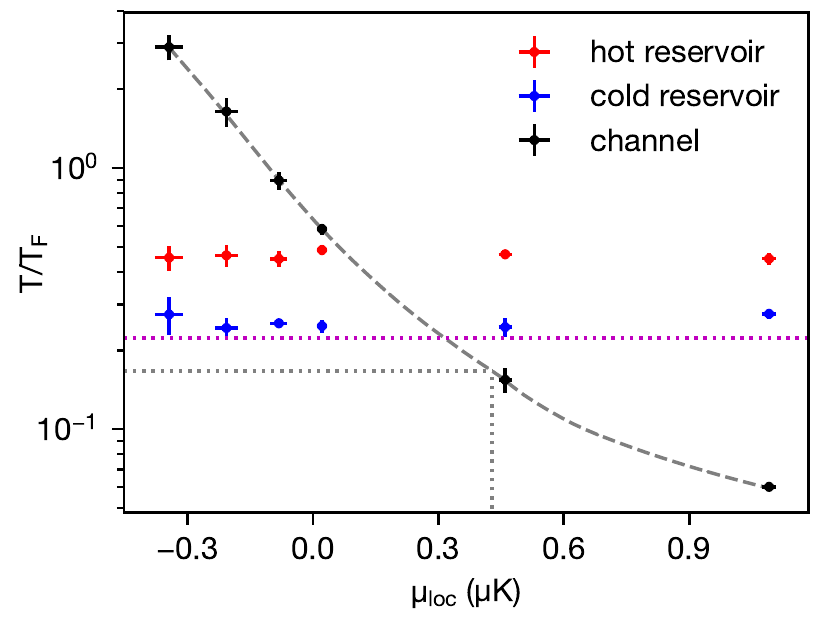}
    \caption{Prepared degeneracy $T/T_F$ inside the hot and cold reservoirs and the channel center versus local chemical potential~$\mu_\text{loc}$. The gray dotted lines locate the normal-to-superfluid transition inside the locally homogeneous channel at $T_c / T_F = 0.167$ and $\mu_\text{loc} = \SI{0.428}{\micro\kelvin}$, respectively. The violet horizontal line indicates the critical degeneracy in the reservoirs. Error bars represent the standard deviation over five repetitions prepared under the same conditions.}
    \label{app_fig:sf_transition}
\end{figure}

The transition is captured by the thermodynamics of the gas and characterized by the degeneracy $T / T_F$ as presented in Fig.~\ref{app_fig:sf_transition} versus the local chemical potential. At the channel center (black points), the gas is in a hydrodynamic regime thanks to the strong interparticle interactions [Appendix~\ref{app_sec:boltzmann}]; thus, it is represented by the local equilibrium described by the average temperature $\bar{T} = \SI{172}{nK}$ and the local chemical potential $\mu_\text{loc}$ [Appendix~\ref{app_sec:eff_pot}]. Therefore, with increasing chemical potential, the degeneracy increases as stated by the thermodynamics of a homogeneous unitary Fermi gas [Appendix~\ref{app_sec:homogeneous_unitary_gas}] and eventually crosses the critical value $T_c / T_F = 0.167(13)$ at $\mu_\text{loc} = \SI{0.428}{\micro\kelvin}$ (gray dotted lines) \cite{Ku2012}.

The hot (red points) and cold (blue points) reservoirs are not affected by the gate potential, and hence, their degeneracy remains constant. They are defined by a half-harmonic potential where the densest point at the center specifies the transition at a critical value $T_c / T_F = 0.223$ (horizontal, violet dotted line). Note that the value depends on the geometry via the Fermi temperature $T_F$. In summary, the reservoirs are both above the critical temperature, while the channel center transitions from a non-condensed to a superfluid state depending on the chemical potential. Thus, in our system, the gas may become superfluid inside the channel.

\section{Boltzmann Approach at Unitarity}
\label{app_sec:boltzmann}

Inside the channel, interatomic collisions may alter the transport regime from ballistic to hydrodynamic depending on how frequently they scatter while crossing the constriction. To assess the regime at unitarity, we follow the Boltzmann approach outlined in Ref.~\cite{Gehm2003_0} and compare the resulting mean free path with the length of the channel. First, we focus on the center, where the gas is homogeneous, and then extend the argument for the full channel.

\subsection{Scattering Time}

Within Boltzmann theory, the scattering rate $\gamma$ follows from the collision integral and reads \cite{Gehm2003_0}
\begin{multline}
	\gamma N = \frac{m}{\pi^2 \hbar^3} \int \! \mathrm{d}E_1 \mathrm{d}E_2 \mathrm{d}E_3 \mathrm{d}E_4 \,
	\delta (E_1 + E_2 - E_3 - E_4) \\
	\times (1 - f_1) (1 - f_2) f_3 f_4 \, H(E, E_m),
\end{multline}
where $N$ denotes the number of particles in each hyperfine state. The integral describes two particles with energies $E_3$ and $E_4$ that exchange energy during the collision and leave with energies $E_1$ and $E_2$. As the collision is elastic, the total particle energy $E = E_1 + E_2 = E_3 + E_4$ is conserved, as visible from the delta function. The product of the Fermi-Dirac distributions $f_i = 1 / (1 + \mathrm{e}^{\epsilon_i})$ with $\epsilon_i = (E_i - \mu) / k_B T$ ensures that the initial states are occupied and the final ones unoccupied, as dictated by Pauli's principle. The factor $H$ is the integrated density of states weighted by the collision cross section $\sigma (k)$. In our situation, the gas is locally homogeneous at the center, which leads to
\begin{equation}
	H(E, E_m) = \frac{2 \pi m V}{(2 \pi \hbar)^3}
	\int_{P_{-}}^{P_{+}} \! \mathrm{d}P \, \sigma (k),
\end{equation}
with the volume $V$ and the integral over total momentum $P$ whose bounds are $P_\pm = \sqrt{2 m (E - E_m)} \pm \sqrt{2 m E_m}$, with the minimal energy $E_m = \min(E_1, E_2, E_3, E_4)$. At unitarity, the collision cross section is $\sigma (k) = 4 \pi / k^2$, expressed with the relative wave vector $k$ given by $4 (\hbar k)^2 = P_{+}^2 + P_{-}^2 - P^2$.

We rewrite the integral by realizing that it is symmetric under the exchange of all four energies. Hence, we can choose $E_1$ to be the lowest energy $E_m$ and include the other cases by a factor four. To ensure that $E_1$ is the smallest and that the total energy is conserved, we transform the variables $E_i$ to the non-negative $x_i$ and normalize with the Fermi energy $E_F = \hbar^2 / (2 m) (6 \pi^2 n)^{2/3}$, with the particle density $n = N / V$, 
\begin{align}
	E_1 / E_F &= x_1, \\
	E_2 / E_F &= x_1 + x_2 + x_3, \\
	E_3 / E_F &= x_1 + x_2, \\
	E_4 / E_F &= x_1 + x_3.
\end{align}

The resulting scattering rate $\gamma = \gamma_0 I (T/T_F)$ consists of a prefactor $\gamma_0$ and a dimensionless integral $I (T/T_F)$. For the prefactor, we choose the classical collision rate in a homogeneous gas at temperature $T_F$, 
\begin{equation}
	\gamma_0 = n \cdot \sqrt{2} \bar{v} \cdot \sigma(k_F) =
	\frac{8 \sqrt{2}}{3 \pi^{3/2}} \frac{E_F}{\hbar},
	\label{app_eq:gamma_0}
\end{equation}
with the average relative velocity $\sqrt{2} \bar{v}$ between two particles expressed through their individual mean speed $\bar{v} = \sqrt{8 E_F / \pi m}$. The dimensionless integral reads
\begin{multline}
	I \left( \frac{T}{T_F} \right) = \frac{9 \sqrt{\pi}}{4}
	\int_0^\infty \! \mathrm{d}x_1 \mathrm{d}x_2 \mathrm{d}x_3 \,
	f(x_1 + x_2) f(x_1 + x_3) \\ \times [1 - f(x_1)] [1 - f(x_1 + x_2 + x_3)] \\
	\times F(2 x_1 + x_2 + x_3, x_1),
\end{multline}
with $f(x) = 1 / (1 + \mathrm{e}^{\xi})$ and $\xi = (T_F / T) (x - \mu / E_F)$, where the normalized temperature $T / T_F$ and chemical potential $\mu / E_F$ are related as in Appendix~\ref{app_sec:homogeneous_unitary_gas}. The function $F(x, x_m)$ depends on the normalized total energy $x = E / E_F$ and minimal energy $x_m = x_1$ as
\begin{equation}
	F(x, x_m) = \frac{1}{\sqrt{x}}
	\log \left( \frac{2 + x/x_m + 2 \sqrt{2 x / x_m}}{2 + x/x_m - 2 \sqrt{2 x / x_m}} \right).
\end{equation}

This form is unsuited for numerical evaluation as it contains a divergence at $x = 0$, which we lift with the substitution $w = x_1$, $y = (x_2 + x_3) / 2 x_1$, and $z = (x_3 - x_2) / 2 x_1$, leading to the final result
\begin{multline}
	I \left( \frac{T}{T_F} \right) = \frac{9}{2} \sqrt{\frac{\pi}{2}}
	\int_0^\infty \! \mathrm{d}y \, \log \left( \frac{1 + y/2 + \sqrt{1 + y}}{1 + y/2 - \sqrt{1 + y}} \right) \\
	\times \int_0^\infty \! \mathrm{d}w \, [1 - f(w)] [1 - f(w (1 + 2 y))] \sqrt{\frac{w^3}{1 + y}} \\
	\times \int_{-y}^y \! \mathrm{d}z \, f(w (1 + y - z)) f(w (1 + y + z)).
\end{multline}

Instead of scattering rates, we use their inverse, the average time between interparticle collisions $\tau = \tau_0 / I (T/T_F)$.

\subsection{Mean Free Path}

Based on the average scattering time calculated in the previous section, we estimate the mean free path $l_\text{mfp} = \tau v_F$, which quantifies the distance a particle travels between collisions with velocity $v_F = \sqrt{2 E_F / m}$. For the estimation, we need the local degeneracy $\bar{T} / T_F$ and the Fermi energy $E_F$ at the center. They both follow from the temperature $\bar{T}$ and chemical potential $\bar{\mu}$ via the local chemical potential $\mu_\text{loc}$, as detailed in Appendix~\ref{app_sec:eff_pot}, and the thermodynamics of a homogeneous unitary Fermi gas given in Appendix~\ref{app_sec:homogeneous_unitary_gas}.

\subsection{Transport Regime}

\begin{figure}
    \includegraphics{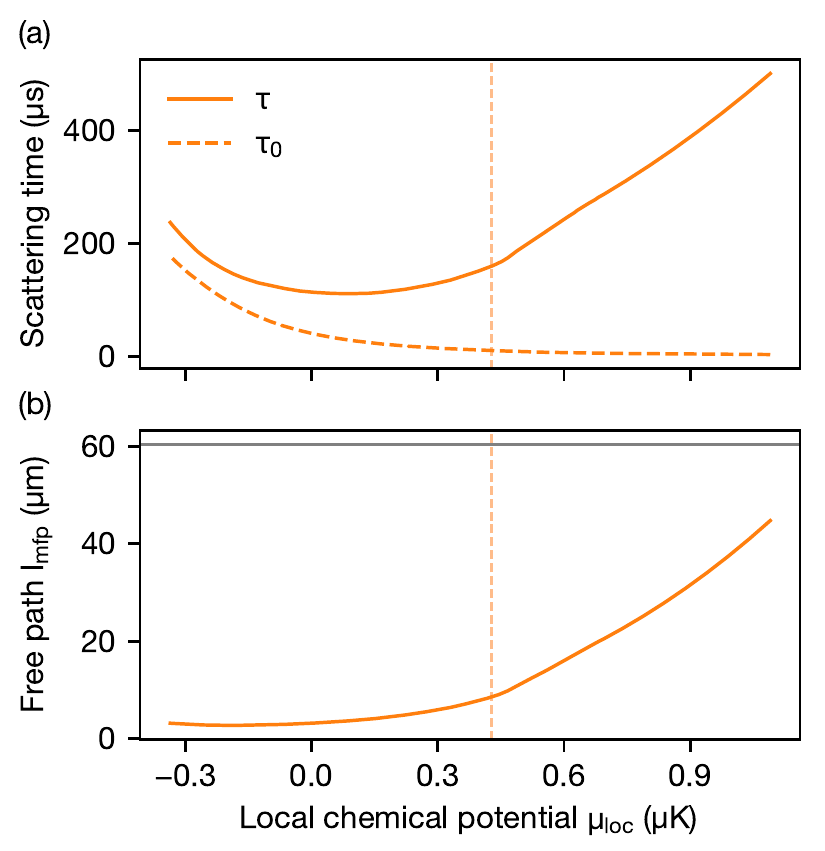}
    \caption{Boltzmann approach for a homogeneous unitary Fermi gas at the channel center. (a) Interparticle scattering time in classical (dashed curve) and quantum (solid line) regimes and (b) mean free path versus local chemical potential. The horizontal line indicates the channel length $2 w_{cy}$ in terms of the waist $w_{cy}$ of the laser beam which creates the channel. The superfluid transition (vertical dashed line) is estimated in Appendix~\ref{app_sec:superfluid_transition}. The predictions are valid in the normal regime above the critical region. In the critical region, we expect the real values to be lower (see text), and in the superfluid regime, the theory fails.}
    \label{app_fig:mean_free_path}
\end{figure}

To discuss the transport regime, we first focus on the scattering time and mean free path at the channel center [Fig.~\ref{app_fig:mean_free_path}]. The classical scattering time~$\tau_0$ versus local chemical potential (dashed line) monotonically decreases, which indicates more frequent collisions. Although the cross section $\sigma(k_F) \sim 1/E_F$ reduces as particles collide at higher relative velocities, the simultaneous increase in density $n \sim E_F^{3/2}$ and mean particle speed $\bar{v} \sim E_F^{1/2}$ dominates [compare with Eq.~\eqref{app_eq:gamma_0}]. In the quantum case (solid line), the time~$\tau$ behaves the same at small chemical potentials. In contrast, at larger values, the gas is more degenerate, and Pauli's principle limits collisions as it requires the final states to be empty. Eventually, this effect dominates and leads to a longer time between collisions. The local minimum in scattering time appears less pronounced in the mean free path as the Fermi velocity monotonically increases [Fig.~\extref[(b)]{app_fig:mean_free_path}].

The predictions based on Boltzmann's approach are valid in the normal Fermi liquid phase located above the critical and superfluid regimes. In the critical region, pairing correlations are relevant, and they modify the scattering rate. In a harmonic trap, they were found to almost compensate Pauli blocking, giving rates that follow the classical prediction \cite{Riedl2008}. Hence, in the homogeneous case, we expect that Pauli blocking in the critical region is also reduced by correlations, giving effectively lower scattering times and mean free paths. Above the superfluid transition (vertical dashed line), the mean free path is bounded by Boltzmann's theory and is at most \SI{9}{\micro\meter}, shorter than the channel length of $2 w_{cy} = \SI{60}{\micro\meter}$.

Next, we extend the argument to the entire channel whose variations are captured by the effective potential [Appendix~\ref{app_sec:eff_pot}]. If the gas is locally non-superfluid throughout the constriction, the mean free path is bounded by Boltzmann's theory, which predicts hydrodynamic transport. This is the case above the transition indicated in Fig.~\ref{fig:evolution}, \ref{fig:thermopower}, \ref{fig:lorenz}, and \ref{app_fig:mean_free_path} [Appendix \ref{app_sec:superfluid_transition}]. Furthermore, as a result of frequent interparticle collisions, we also expect the reservoirs to be in the hydrodynamic regime.

The Onsager framework adopted in Sec.~\ref{sec:phenomenological_model} is valid in the unitary regime but differs in two aspects from a standard hydrodynamic description. (i) The presented Onsager picture relates the flow to global differences in intensive quantities between the reservoirs, while in hydrodynamic models, local gradients are relevant. (ii) The intensive quantities are typically pressure and temperature in the hydrodynamic formulation, while our treatment is based on chemical potential and temperature. The latter combination naturally leads to a symmetric matrix of transport coefficients thanks to Onsager's reciprocal relation [Eq.~\eqref{eq:phenomenological_model}]. 

However, setting aside the symmetry, the choice of pressure and temperature leads to an equivalent description by transforming the intensive quantities via the thermodynamics of the gas. Namely, with the help of the Gibbs-Duhem relation, the difference in pressure, 
\begin{equation}
	\Delta p = s \Delta\mu + n \Delta T, 
\end{equation}
at the interface between the reservoirs is directly related to the corresponding differences in chemical potential~$\Delta \mu$ and temperature $\Delta T$. The coefficients are given by the homogeneous densities of entropy~$s$ and particles~$n$ at the trap center, evaluated at the average chemical potential~$\bar{\mu}$ and temperature~$\bar{T}$. Then, in the reservoir-dominated regime, the flow is produced by a difference in pressure, which is analogous to our interpretation with a difference in chemical potential. Understanding the channel-dominated regime, where the particles flow against the differences $\Delta p$ and $\Delta \mu$, requires a hydrodynamic model of the channel, in particular, of its mode structure. Instead, we estimate the thermopower~$\alpha_c$ of the channel by the dilatation coefficient at the channel center, which offers an interpretation of the interaction-assisted reversal [Sec.~\ref{sec:thermopower}].

\section{Effective Potential}
\label{app_sec:eff_pot}

In two-terminal setups, particles move along one direction and are confined in the others. As a result of the confinement, the transverse motion is quantized into different modes labeled by quantum numbers. As long as the confinement varies adiabatically through the channel, particles remain in the same mode, which is called the adiabatic approximation. Then, the confinement energy acts as an additional potential as it is invested in the transverse direction and is missing for the longitudinal motion. In the following, we present the effective potential of our channel, which is useful to calculate the local chemical potential at the center and to deduce transport parameters [Appendix~\ref{app_sec:landauer}].

Harmonically approximating our transverse confinement ($x$, $z$) at each longitudinal position ($y$) leads to the potential
\begin{equation}
	V (x, y, z) = \frac{1}{2} m \omega_x^2 (y) x^2 + \frac{1}{2} m \omega_z^2 (y) z^2 + V_g (y).
\end{equation}

The harmonic frequencies $\omega_x (y)$ and $\omega_z (y)$ include a nearly constant contribution from the dipole trap due to the long Rayleigh length of \SI{20}{\milli\meter} and a spatially varying one from the gate potential and the one creating the channel, respectively. They are given by
\begin{align}
	\omega_x^2 (y) &= \omega_{tx}^2 + \omega_{gx}^2 \mathrm{e}^{- 2 y^2 / w_{gy}^2}, \\
	\omega_z^2 (y) &= \omega_{tz}^2 + \omega_{cz}^2 \mathrm{e}^{- 2 y^2 / w_{cy}^2}, 
\end{align}
with the dipole trap frequencies $\omega_{tx/tz} = 2 \pi \, \nu_{tx/tz}$, and the frequencies at the center created by the gate $\omega_{gx}^2 = -4 V_g / m w_{gx}^2$ and channel beam $\omega_{cz}$. Their waists in the longitudinal direction are $w_{gy}$ of \SI{8.6}{\micro\meter} for the repulsive gate beam, \SI{33.5}{\micro\meter} for the attractive gate beam, and $w_{cy} = \SI{30.2}{\micro\meter}$ for the channel beam. Besides modifying the trapping frequency, the gate beam creates an additional potential 
\begin{equation}
	V_g (y) = V_g \mathrm{e}^{- 2 y^2 / w_{gy}^2}.
\end{equation}

As at each longitudinal position~$y$, the transverse potential is quadratic and shifted by the energy $V_g (y)$, the eigenenergies are
\begin{equation}
	E_\mathbf{n} (y) = \hbar \omega_x (y) (n_x + 1/2) + \hbar \omega_z (y) (n_z + 1/2) + V_g(y), 
	\label{app_eq:eigenenergies}
\end{equation}
with the quantum number $\mathbf{n} = (n_x, n_z)$. Figure~\ref{app_fig:eff_pot} displays the eigenenergies in the transport direction for different gate strengths $V_g$.

\subsection{Local Chemical Potential}

To characterize the gas at the channel center, we use the local density approximation with the effective potential $V_\text{eff} (V_g) = E_\mathbf{0} (0)$ in the ground state, which results in a local chemical potential $\mu_\text{loc} = \bar{\mu} - V_\text{eff} (V_g)$, with $V_\text{eff} (V_g) = \hbar (\omega_x (0) + \omega_z (0)) / 2 + V_g$. Note that $\omega_x (0)$ implicitly depends on $V_g$ in a square-root fashion. The local chemical potential is used to display transport coefficients [Figs.~\ref{fig:thermopower} and \ref{fig:lorenz}], to locate the superfluid transition [Appendix~\ref{app_sec:superfluid_transition}], and to discuss its scattering properties [Appendix~\ref{app_sec:boltzmann}].

\subsection{Transport Function~$\Phi (E)$}

\begin{figure*}
    \includegraphics{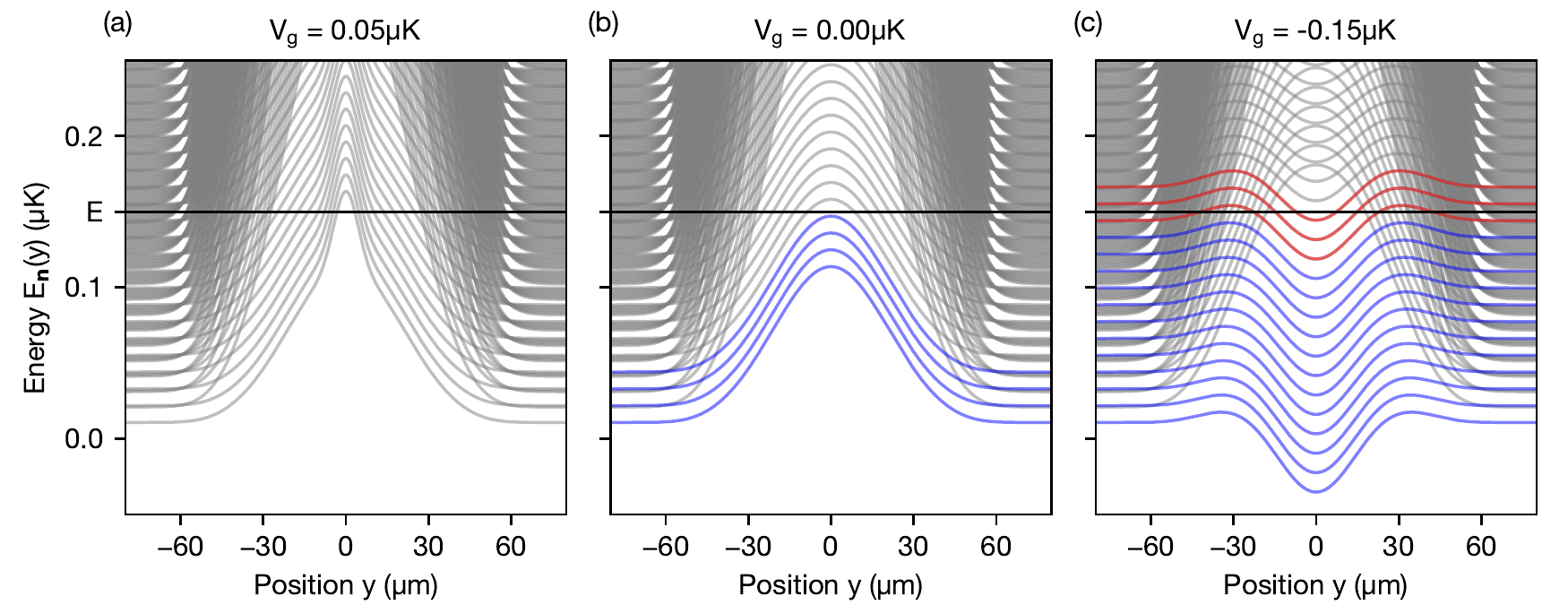}
    \caption{Eigenenergies~$E_\mathbf{n}$ with transverse mode $\mathbf{n}$ versus longitudinal position $y$ (a) with repulsive, (b) without and (c) with attractive gate potential $V_g$. Modes available throughout the channel at energy $E$ (horizontal line) are indicated in blue and contribute to transport. The ones in red are accessible close to the center but blocked away from it and the remaining modes are plotted in gray.}
    \label{app_fig:eff_pot}
\end{figure*}

From the reservoirs to the center, the channel narrows and tends to increase the mode energies, visible in Fig.~\ref{app_fig:eff_pot} in the absence of the gate ($V_g = 0$). Here, a particle at energy $E$ (horizontal line) can cross the channel in any mode indicated in blue. An additional repulsive potential ($V_g > 0$) pushes them up and further peaks them at the center within the size of the beam. In the attractive case ($V_g < 0$), the energies are pulled down, and some modes might be energetically allowed at the center while away from the center, the mode energies are above the particle energy (red lines). Only the modes whose energies are below the particle energy throughout the channel are relevant for transport. Their number is counted with the transport function $\Phi (E)$ directly from the effective potential and is used in Appendix~\ref{app_sec:landauer} to calculate transport parameters.

\section{Landauer-Büttiker Theory}
\label{app_sec:landauer}

In this Appendix, we detail how we model the transport parameters in the non-interacting situation. In the Landauer framework, particles come from the reservoirs following the corresponding Fermi-Dirac distribution, and they cross the channel with a transmission probability. In linear response, the transport coefficients are evaluated at the mean chemical potential $\bar{\mu}$ and temperature $\bar{T}$ imposed by the reservoirs. They read as follows \cite{Grenier2016, Butcher1990, Sivan1986}: 
\begin{align}
	G &= \frac{1}{h} \int_{-\infty}^{+\infty} \! \Phi (E)
	\left( - \frac{\partial f (\epsilon)}{ \partial E} \right) \,\mathrm{d}E, \label{app_eq:landauer_G} \\
	G \alpha_c &= \frac{1}{h \bar{T}} \int_{-\infty}^{+\infty} \! \Phi (E) (E - \bar{\mu})
	\left( - \frac{\partial f (\epsilon)}{ \partial E} \right) \,\mathrm{d}E, \label{app_eq:landauer_G_alpha_c} \\
	G (L + \alpha_c^2) &= \frac{1}{h \bar{T}^2} \int_{-\infty}^{+\infty} \! \Phi (E) (E - \bar{\mu})^2
	\left( - \frac{\partial f (\epsilon)}{ \partial E} \right) \,\mathrm{d}E, \label{app_eq:landauer_G_L_p_alpha_c2}
\end{align}
with the Fermi-Dirac distribution $f (\epsilon) = 1 / (1 + \mathrm{e}^\epsilon)$ and the normalized particle energy $\epsilon = (E - \bar{\mu}) / k_B \bar{T}$.

In the classical regime, the transmission $\Phi (E)$ through the channel reduces to the number of transverse modes below energy $E$. We count their number directly from the effective potential discussed in Appendix~\ref{app_sec:eff_pot}. Then, by numerically evaluating the Landauer integrals, the transport coefficients follow and are indicated in Figs~\ref{fig:thermopower} and \ref{fig:lorenz}.

\subsection{Benchmarking}

To benchmark the method, we compare it with the measured conductance for a non-interacting gas as shown in Fig.~\extref[(a)]{fig:thermopower} and find good agreement. Note that counting available modes at the channel center leads to a wrong prediction that increases roughly quadratically with local chemical potential, in contrast to the observed linear behavior.

\subsection{Validity of Linear Response}

\begin{figure}
    \includegraphics{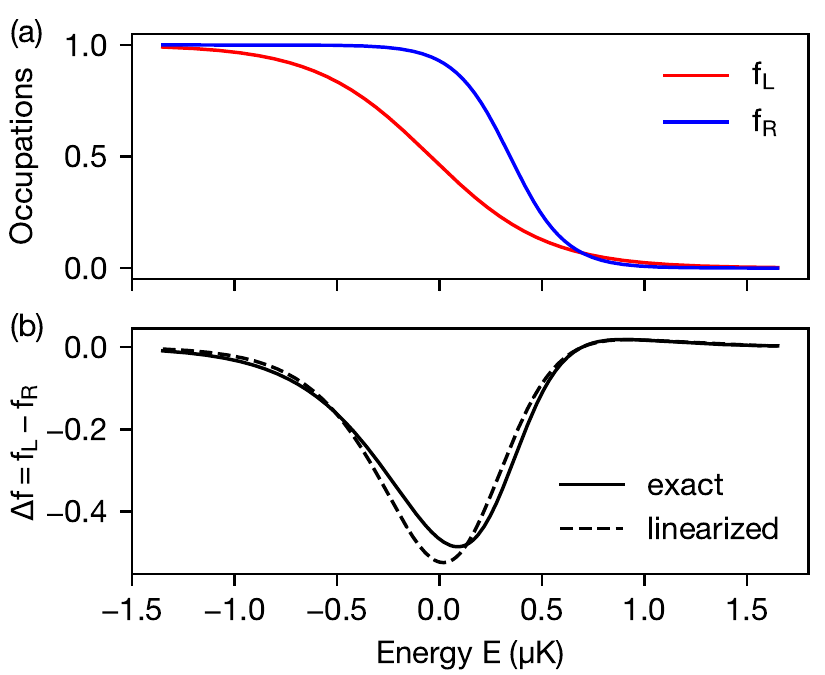}
    \caption{Validity of linear response. (a) Fermi-Dirac distributions in both reservoirs as prepared for the non-interacting measurement and (b) their exact (solid) and linearized (dashed) difference. The numbers used are summarized in Table~\ref{app_tab:exp_conditions}.}
    \label{app_fig:occupation}
\end{figure}

In Landauer theory, particle and entropy currents are expressed with the difference $\Delta f (E) = f_L (E) - f_R (E)$ between the Fermi-Dirac distributions. To arrive at the linear response form [Eqs.~\eqref{app_eq:landauer_G}--\eqref{app_eq:landauer_G_L_p_alpha_c2}], the difference is expanded around the average chemical potential $\bar{\mu}$ and temperature~$\bar{T}$, and we obtain 
\begin{equation}
	\Delta f (E) \simeq - \frac{\partial f (\epsilon)}{ \partial E}
	\left( \Delta \mu + \epsilon k_B \Delta T \right), 
\end{equation}
with the derivative $- \partial f (\epsilon) / \partial E = 1 / (2 k_B \bar{T} [1 + \cosh(\epsilon)])$. The distributions and the exact and linearized differences are displayed in Fig.~\ref{app_fig:occupation} for the conditions of the non-interacting measurement. Visibly, linearization only introduces minor deviations, justifying the approximation.

\end{document}